\documentclass[]{jfm}

\usepackage{amsmath}
\usepackage{comment}
\usepackage[normalem]{ulem}
\usepackage{graphicx}
\usepackage{epsfig}
\usepackage{amsfonts}
\usepackage{caption}
\usepackage{subcaption}
\usepackage{natbib}
\usepackage[dvipsnames]{xcolor}
\usepackage{hyperref}
\hypersetup{
    colorlinks = true,
    urlcolor   = blue,
    citecolor = black,
    linkcolor = black
}

\newcommand{\pd}{\partial}
\newcommand{\wh}{\widehat}
\newcommand{\avg}[1]{\langle{#1}\rangle}
\newcommand{\uhk}[1]{\wh{\bf u}_{#1}(\bold k)}

\newcommand{\chk}[1]{\textcolor{black}{#1}}
\newcommand{\vd}[1]{\textcolor{black}{#1}}

\newcommand{\vbj}[1]{\textcolor{black}{#1}}
\title{Relaxation and statistical equilibria in generalised two-dimensional flows}

\author{
Vibhuti Bhushan Jha\aff{1,2}\corresp{\email{vibhutibjha@gmail.com}},
Kannabiran Seshasayanan\aff{2}\corresp{\email{kanna@iitm.ac.in}}
\and 
Vassilios Dallas\aff{3}\corresp{\email{vassilios.dallas@gmail.com}}
}

\affiliation{
\aff{1}Space Applications Centre, Indian Space Research Organisation, Ahmedabad, Gujarat, India
\aff{2}Department of Applied Mechanics and Biomedical Engineering, Indian Institute of Technology Madras, Chennai, India
\aff{3}Environmental Research Laboratory, National Centre for Scientific Research ``Demokritos'', 15341 Athens, Greece
}

\begin{document}
\maketitle

\begin{abstract}
We study relaxation toward statistical equilibrium states of inviscid generalised two-dimensional flows, where the generalised vorticity $q$ is related to the streamfunction $\psi$ via $q=(-\nabla^2)^{\frac{\alpha}{2}}\psi$, with the parameter $\alpha$ controlling the strength of the nonlinear interactions. The equilibrium solutions exhibit an $\alpha \mapsto -\alpha$ symmetry, under which generalised energy $E_G$ and enstrophy $\Omega_G$ are interchanged. 
For initial conditions that produce condensates, we find long-lived quasi-equilibrium states far from the thermalised solutions we derive using canonical ensemble theory.
Using numerical simulations we find that in the limit of vanishing nonlinearity, as $\alpha \to 0$, 
the time required for partial thermalisation $\tau_{th}$ scales like $1/\alpha$. So, the relaxation of the system toward equilibrium becomes increasingly slow as the system approaches the weakly nonlinear limit. This behaviour is also captured by a reduced model we derive using multiple scale asymptotics. 
These findings highlight the role of nonlinearity in controlling the relaxation toward equilibrium \vd{and that the inherent symmetry of the statistical equilibria determines the direction of the turbulent cascades.
}
\end{abstract}


%
\section{Introduction}

The emergence of coherent structures in two-dimensional (2D) turbulence has long motivated the use of statistical mechanics as a theoretical framework, beginning with Onsager’s pioneering theory of point vortices \citep{onsager1949statistical}. 
Persistent vortices such as Jupiter’s Great Red Spot, the ubiquity of inverse energy cascades in the atmosphere and ocean, and the organisation of planetary flows into zonal jets, all highlight the tendency of 
geophysical systems to self-organise \citep{byrne2013height,king2015upscale,shao2023physical,balwada2022direct,ScottWang2005,young2017forward}.
The large-scale dynamics of such systems are often modelled using the incompressible 2D Euler equations, which provide a natural setting for equilibrium statistical mechanics \citep{miller1992statistical,bouchet2012statistical}. In this framework, the large-scale coherent flow is interpreted as an emergent equilibrium state constrained by the conserved quantities of the system. Several studies have extended this approach to geophysically relevant flows \citep{salmon1998lectures,majda2006nonlinear,bourouiba2008model,TeitelbaumMininni2012,weichman2022statistical}.

\vd{For the truncated two-dimensional (2D) Euler equations, \citet{kraichnan1975statistical} derived the long time statistical equilibrium solutions using the canonical and microcanonical ensemble theory. 
For the truncated three-dimensional (3D) Euler equations, \citet{kraichnan1973helical} showed that there are no negative temperature states.
During relaxation toward equilibrium, subsets of Fourier modes thermalise and act as effective microstates, generating an 
effective viscosity effect on the modes that have not yet thermalised \citep{cichowlas2005effective,bos2006dynamics}.
Moreover, studies have shown
the truncated 3D Euler equations 
to reproduce aspects of Navier-Stokes dynamics at large scales \citep{krstulovic2009cascades}, and conversely, the large-scale flow of 3D Navier-Stokes equations \chk{also} exhibit features characteristic of the truncated 3D Euler dynamics \citep{Dallasetal2015,alexakis2019thermal}. 
}

Understanding how a system relaxes toward equilibrium is a long-standing challenge in statistical mechanics. A central factor that controls relaxation toward equilibrium in such systems is the strength of the nonlinear interactions. The classical Fermi-Pasta-Ulam (FPU) problem exemplifies how weak nonlinearity and near-integrability can significantly delay relaxation, as energy is transferred only through weak wave-wave interactions \citep{onorato2015route,onorato2023wave}.
In the weakly nonlinear regime, the timescale for equilibration can increase dramatically, resulting in slow or even incomplete thermalisation.

The generalised 2D flows provide a systematic way to tune the effective nonlinearity of the governing equations through a parameter $\alpha$, which links the generalised vorticity $q$ and the streamfunction $\psi$ via $q = (-\nabla^2)^{\alpha/2}\psi$ \citep{pierrehumbert1994spectra}. This formulation unifies a wide class of geophysical models: the barotropic vorticity equation for $\alpha=2$ \citep{Tabeling2002,BoffettaEcke2012}, the surface quasi-geostrophic (SQG) equation for $\alpha=1$ \citep{held1995surface,Lapeyre2017}, and a rescaled shallow-water quasi-geostrophic model for $\alpha=-2$ \citep{LarichevMcWilliams1991,Smithetal2002}.
Recent work by \citet{jha2025cascades} studied the \vd{transition in the direction of cascades of the inviscid conserved quantities as $\alpha$ varies} in the forced-dissipative system. Positive values of $\alpha$ showed organisation of large scale vortices, while negative values of $\alpha$ showed the formation of filamentary structures reminiscent of 3D isotropic turbulence.

In the limit $\alpha \to 0$, the nonlinear terms in the generalised model become progressively weaker and ultimately vanish, yielding a laminar flow state \citep{jha2025cascades}. This controlled reduction of nonlinearity naturally raises questions about how the dynamics and the relaxation to equilibrium of the inviscid system depend on $\alpha$, \vd{reminiscent in a way of the delayed thermalisation} seen in the FPU model. Recent works on thermalisation in truncated 2D \vd{systems} showed that for negative temperature states, where the energy is concentrated at small wavenumbers, \vd{the flows evolve} to a partially thermalised state at long times \citep{venaille2015violent,Agouaetal2025}. Here, we focus on 
the long time thermalisation states \vd{of the generalised 2D model}, and on understanding the dependence on the $\alpha$ parameter, 
\vd{as well as the limit of $\alpha \rightarrow 0$, where the nonlinearity tends to vanish}. 

The objectives of this study are threefold. 
First, to derive the statistical equilibrium states of the inviscid generalised 2D flows for different values of $\alpha$ using the canonical ensemble theory. 
Second, to study the symmetry of the thermalised solutions for positive and negative $\alpha$.
Third, to quantify how the time for thermalisation 
and spectral organisation depends on $\alpha$, particularly in the weakly nonlinear limit $\alpha \to 0$. 
\chk{The paper is organised as follows. In Section \ref{sec:theory} we formulate the problem, we derive the statistical equilibrium solutions 
and we discuss their $\alpha \mapsto -\alpha$ symmetry. Moreover, using multiple scale asymptotics, we derive a reduced model for the weakly nonlinear limit $\alpha \to 0$. In Section \ref{sec:results}, we present the numerical results on the relaxation toward equilibrium and the dependence of the time to reach thermalisation on the parameter $\alpha$. Finally, we summarise our concluding remarks in Section \ref{sec:conc}.}

\section{Problem formulation and \vd{statistical equilibrium} theory}
\label{sec:theory}

Following \cite{pierrehumbert1994spectra}, \vd{we} consider the 2D advection equation for the generalised vorticity $q$ of an inviscid fluid,
\begin{equation}
	\partial_tq + J(\psi,q) = 0,
	\label{eq:gmodel}
\end{equation}
where $\psi$ is the streamfunction which is related to $q$ via 
\begin{equation}
	q = (-\nabla^2)^{\frac{\alpha}{2}} \psi,
	\label{eq:alpha}
\end{equation}
with $\alpha$ being a parameter that generalises the vorticity relationship $q = -\nabla^2 \psi$ 
of 2D fluid dynamics. The divergence-free fluid velocity field is
given by 
${\bf u} = \nabla \times(\psi \hat{\bf z})$.
Equations \eqref{eq:gmodel} and \eqref{eq:alpha} have two quadratic conserved quantities, namely, 
\begin{equation}
  E_G = \avg{\psi q}, \quad \Omega_G = \avg{q^2},
  \label{eq:invariants}
\end{equation}
where $E_G$ is the generalised energy, $\Omega_G$ is the generalised enstrophy 
and the angular brackets $\avg{.}$ denote spatiotemporal averaging.
In Fourier space, relationships \eqref{eq:alpha} and \eqref{eq:invariants} become, respectively 
\begin{align}
  \widehat{q}({\bf k}) &= k^{\alpha} \widehat{\psi}({\bf k}),
   \label{eq:qk} \\
  	   E_G &= \sum_{\bf k} k^{\alpha} |\wh{\psi}(\bold k)|^2 
	    	 	= \sum_{\bf k} k^{\alpha-2} |\uhk{}|^2, 
    \label{eq:inv1} \\
  \Omega_G &= \sum_{\bf k} k^{2\alpha}|\wh{\psi}(\bold k)|^2
		        = \sum_{\bf k} k^{2\alpha-2}|\uhk{}|^2,
    \label{eq:inv2}
\end{align}
where $\,\widehat{.}\,$ denotes Fourier components and $k = |{\bf k}| = \sqrt{k_x^2+k_y^2}$ is the isotropic wavenumber. 

To derive the equilibrium states, we employ a canonical ensemble approach \citep{lee1952,Kraichnan1967,salmon1998lectures}, where the probability density of a given state is determined by the invariants \eqref{eq:inv1} and \eqref{eq:inv2} of the system and is given by
\begin{align}
  \mathcal{P} &= \frac{1}{\mathcal{Z}} \exp(-\beta E_G -\gamma \Omega_G) 
    \label{eq:pdf} \\
 \text{with \;} \mathcal{Z} &= \int_\Gamma \exp(-\beta E_G -\gamma \Omega_G) \; d\Gamma,
   \label{eq:Z}
\end{align}
where $\beta$ and $\gamma$ play the role of inverse temperatures,
$\mathcal{Z}$ is the partition function, and $d\Gamma$ is the phase space volume element, which
is given by 
$d\Gamma = \prod_{\bf k} d\Gamma(\bold k) = \prod_{\bf k} d \uhk{R} \, d \uhk{I}$, 
such that $\uhk{} = \uhk{R} +  i \uhk{I}$. 
Substituting definitions \eqref{eq:inv1} and \eqref{eq:inv2} into \eqref{eq:Z}, we get
\begin{equation}
\mathcal{Z} = \int \exp \left[ -\beta \sum_{\bf k} k^{\alpha-2} |\uhk{}|^2 
				   -\gamma \sum_{\bf k} k^{2\alpha-2} |\uhk{}|^2 \right] 
			\prod_{\bf k} d \uhk{R} \, d \uhk{I}.
\label{eq:Z}
\end{equation}
Assuming isotropy, the above partition function can be converted from an integral over the whole phase space into a product over all wavenumbers. \vbj{Taking into account the incompressibility criterion,} for each wavenumber we can write the individual partition function, denoted as $\mathcal{Z}_k$, as a product of two one-dimensional integrals with respect to the absolute values $|\uhk{R}|$ and $|\uhk{I}|$, as follows
\vd{
\begin{align}
\mathcal{Z}_k \propto 
		& \; \int_{0}^{\infty} \exp \left[ -|\uhk{R}|^2 (\beta k^{\alpha-2}+\gamma k^{2\alpha-2}) \right]  \, d |\uhk{R}| \nonumber \\ 
		\times 
		& \; \int_{0}^{\infty} \exp \left[ -|\uhk{I}|^2 (\beta k^{\alpha-2}+\gamma k^{2\alpha-2}) \right] \, d |\uhk{I}|
\label{eq:Zk}
\end{align}
}
The Gaussian integrals above can be simplified using the following expression,
\begin{equation}
\int_{0}^{\infty} \exp \left[ -u^2(\beta k^{\alpha-2}+\gamma k^{2\alpha-2}) \right] \, du 
		    = \sqrt{\frac{\pi}{4(\beta k^{\alpha-2}+\gamma k^{2\alpha-2})}}
\label{eq:Gint}
\end{equation}
where the argument of the exponential must be positive for the integral to converge, which implies $\beta k^{\alpha-2}+\gamma k^{2\alpha-2} > 0$ or simply $\beta + \gamma k^{\alpha} > 0$. So, the partition function can be expressed as a product over individual mode contributions $\mathcal{Z}_k$ as 
\begin{equation}
  \mathcal{Z} = \prod_k \mathcal{Z}_k 
  \quad \text{where} \quad 
  \mathcal{Z}_k \propto \frac{\pi}{(\beta k^{\alpha-2}+\gamma k^{2\alpha-2})}.
  \label{eq:Zk2}
\end{equation}

The isotropic spectra of $E_G (k)$ and $\Omega_G (k)$ can be derived from the partition function using standard thermodynamic relations \citep{kraichnan1975statistical,TeitelbaumMininni2012} as given below,
\begin{align}
	    E_G(k) = -2 \pi k \frac{\partial \ln\mathcal{Z}_k}{\partial \beta}, \quad
  \Omega_G(k) = -2 \pi k \frac{\partial \ln\mathcal{Z}_k}{\partial \gamma}.
  \label{eq:thermo}
\end{align}

Substituting \eqref{eq:Zk2} \vd{into \eqref{eq:thermo}, we get} 

\begin{align}
	  E_G(k) = \frac{2\pi k}{\beta+\gamma k^{\alpha}}, \quad
\Omega_G(k) = \frac{2\pi k^{\alpha+1}}{\beta+\gamma k^{\alpha}}.
\label{eq:equilspec}
\end{align}
These expressions reduce to well known equilibrium states for the 2D Euler ($\alpha=2$) and the SQG model ($\alpha=1$) as have been reported by \cite{kraichnan1975statistical} and \cite{TeitelbaumMininni2012}, respectively.

\subsection{Equilibrium regimes and the \vd{$\alpha \mapsto - \alpha$} symmetry}
\label{sec:regimes} 
Following \cite{Kraichnan1967,kraichnan1975statistical}, the signs of $\beta$, $\gamma$ and the value of \vd{$k_c^\alpha = \Omega_G / E_G$ at $t=0$} determine the behaviour of the equilibrium spectra \eqref{eq:equilspec} distinguished by three regimes 
    \begin{align}
      &\text{Regime I:} &k_{\min}^{\alpha} < k_c^{\alpha} < k_\beta^{\alpha}, \quad &\gamma > 0, \quad -\gamma k_{\min}^{\alpha} < \beta < 0,
      \label{eq:RegI} \\ 
      &\text{Regime II:} &k_\beta^{\alpha} < k_c^{\alpha} < k_\gamma^{\alpha}, \quad &\beta > 0,\quad \gamma > 0, 
      \label{eq:RegII} \\ 
      &\text{Regime III:} &k_\gamma^{\alpha} < k_c^{\alpha} < k_{\max}^{\alpha}, \quad &\beta > 0,\quad -\beta < \gamma k_{\max}^{\alpha} < 0,
        \label{eq:RegIII}
    \end{align}
\vd{where $k_{\min}$ and $k_{\max}$ are the smallest and largest wavenumber of the system, respectively.} 
The generalised energy and enstrophy equipartition states determine \vd{$k_\beta^\alpha = \lim_{\beta \to 0} k_c^{\alpha}$ and $k_\gamma^\alpha = \lim_{\gamma \to 0} k_c^{\alpha}$, respectively.}

\begin{figure}
    \centering
    \includegraphics[width=0.5\linewidth]{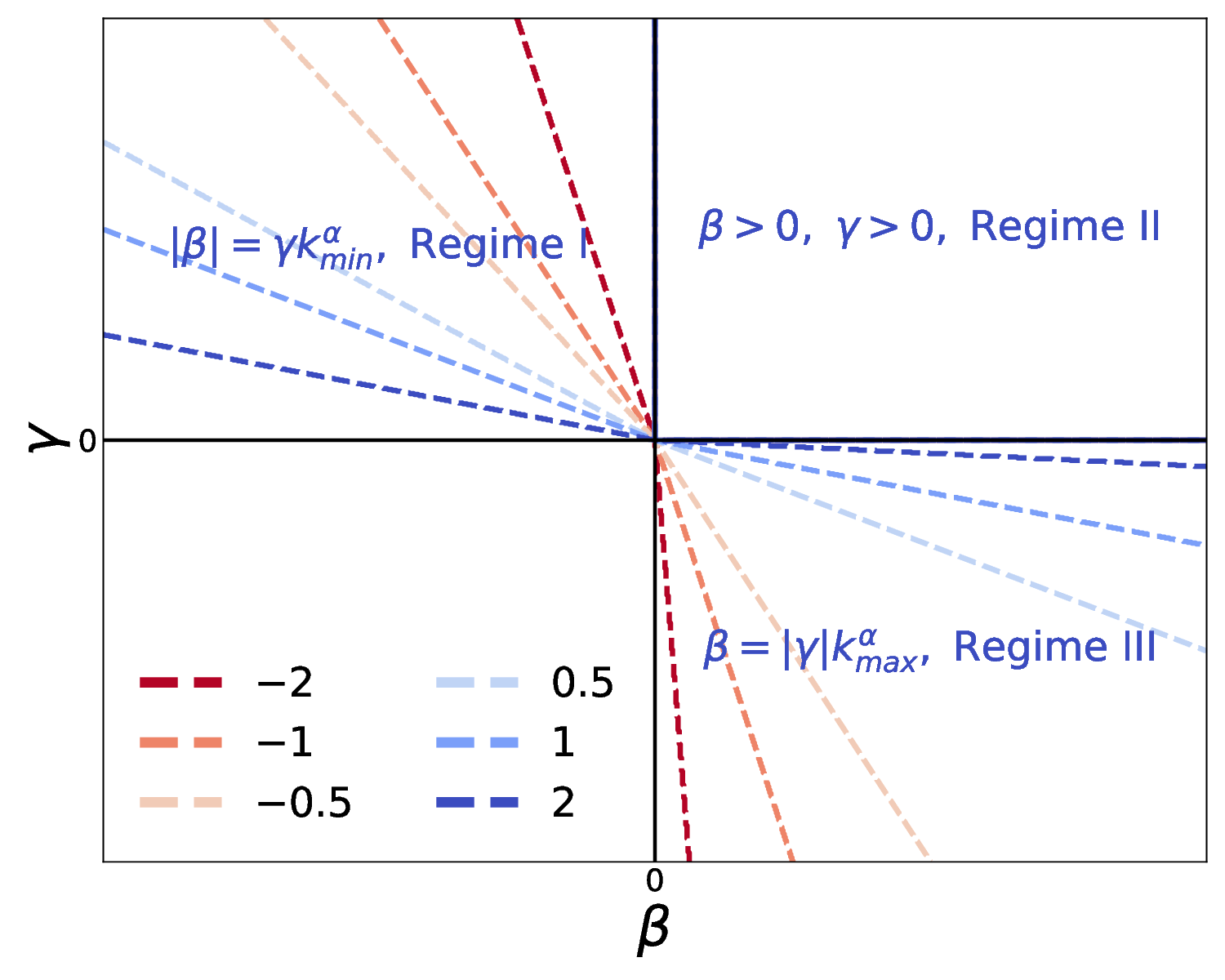}
    \caption{\vd{The three equilibrium regimes on the $\beta$-$\gamma$ parameter space for certain values of $\alpha$. The region with $\beta < 0$, $\gamma > 0$ is Regime I, the region with $\beta, \gamma > 0$ is Regime II and the region with $\beta > 0$, $\gamma < 0$ is Regime III. The lines denote the regions for each value of $\alpha$ shown in the legend.} 
    }
    \label{fig:regimes}
\end{figure}

Figure \ref{fig:regimes} demonstrates the three equilibrium regimes on the $\beta$-$\gamma$ parameter space for certain values of $\alpha$. 
 \vd{Boundaries of the regimes I and III are shown by dashed lines for the different values of $\alpha$.}
Regime I 
is characterised by $\beta < 0$ and $\gamma > 0$ (see Fig. \ref{fig:regimes}). For $\alpha > 0$, this implies that the generalised energy tends to accumulate toward $k_{\min}$, while generalised enstrophy is concentrated towards smaller scales (see Eq. \eqref{eq:equilspec}).
Regime II 
corresponds to $\beta > 0$ and $\gamma > 0$ (see Fig. \ref{fig:regimes}), which is bounded by the energy-equipartition state \vd{($\gamma = 0$)}, $k_c=k_\gamma$, and the enstrophy-equipartition state \vd{($\beta=0$)}, $k_c=k_\beta$, valid for all values of $\alpha$.
Finally, regime III 
is characterised by $\beta > 0$ and $\gamma < 0$ (see Fig. \ref{fig:regimes}).
For $\alpha < 0$, this implies that generalised enstrophy tends to accumulate toward $k_{\min}$, while generalised energy is concentrated towards smaller scales (see Eq. \eqref{eq:equilspec}). 
As \cite{kraichnan1975statistical} noted, the negative temperature states exist only if $k_{\min}$ is non zero. Next, we proceed to show that the $\alpha\to-\alpha$ symmetry helps us understand the behaviour in more detail. 

\vd{For the forced-dissipative} evolution equation of the generalised vorticity $q$, the inverse cascade of $E_G$ and the forward cascade of $\Omega_G$ for $\alpha > 0$, reverse direction when $\alpha < 0$ \citep{jha2025cascades}.
A natural question that arises is if such a behaviour is intrinsic for the ideal system \eqref{eq:gmodel}, \eqref{eq:alpha}. \vd{Taking into account 
Eq. \eqref{eq:alpha}}, the form of the invariants Eqs. \eqref{eq:invariants} and the isotropic spectra Eqs. \eqref{eq:equilspec}, we can deduce that if \vd{$\alpha \mapsto -\alpha$} then the solution is invariant 
\vd{under the interchanges $E_G(k) \leftrightarrow \Omega_G(k)$ and 
$\beta \leftrightarrow \gamma$.}
Since $\beta, \gamma$ are determined by the generalised energy 
and enstrophy, 
the symmetry is also given in terms of the interchange $E_G \leftrightarrow \Omega_G$. The exact expressions for \vd{$E_G$ and $\Omega_G$, are shown in Table \ref{tbl:E_Omega_alpha} of Appendix A} for certain values of $\alpha$, demonstrating the symmetry of \vd{$\alpha \mapsto -\alpha$}. With this result, we can show that the solutions in regime I for $\alpha > 0$ can be mapped to solutions in regime III for $\alpha < 0$. Similarly the points in regime I for $\alpha < 0 $ can be mapped to points in regime III for $\alpha > 0$, \vd{while solutions in regime II are mapped to other points in regime II.}

\begin{figure}
	\centering
	\begin{subfigure}[h]{0.49\textwidth}		
		\includegraphics[width=\linewidth]{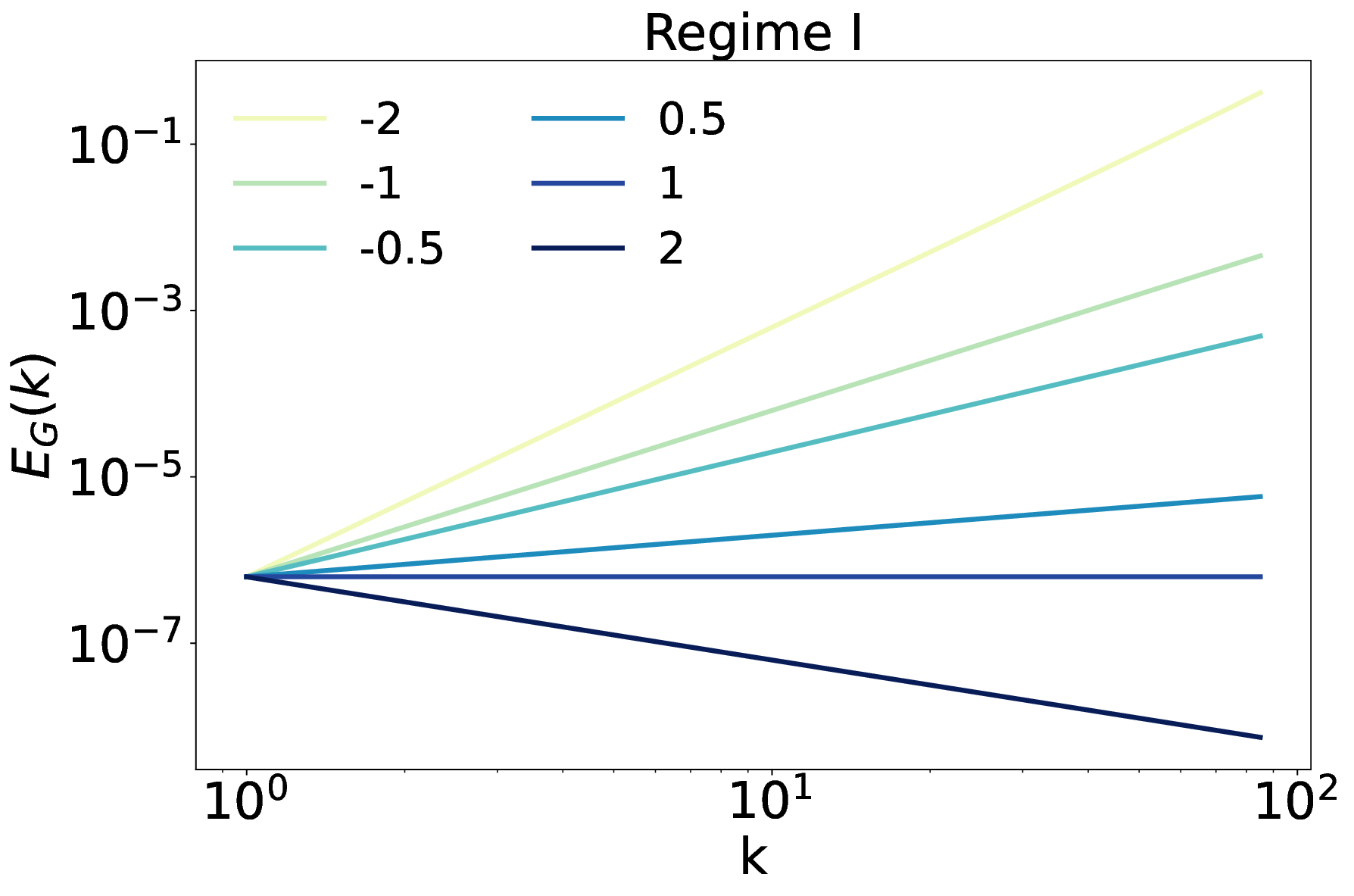}
		\caption{\label{fig:regimes1}}
	\end{subfigure}
	\begin{subfigure}[h]{0.49\textwidth}		
		\includegraphics[clip,width=\linewidth]{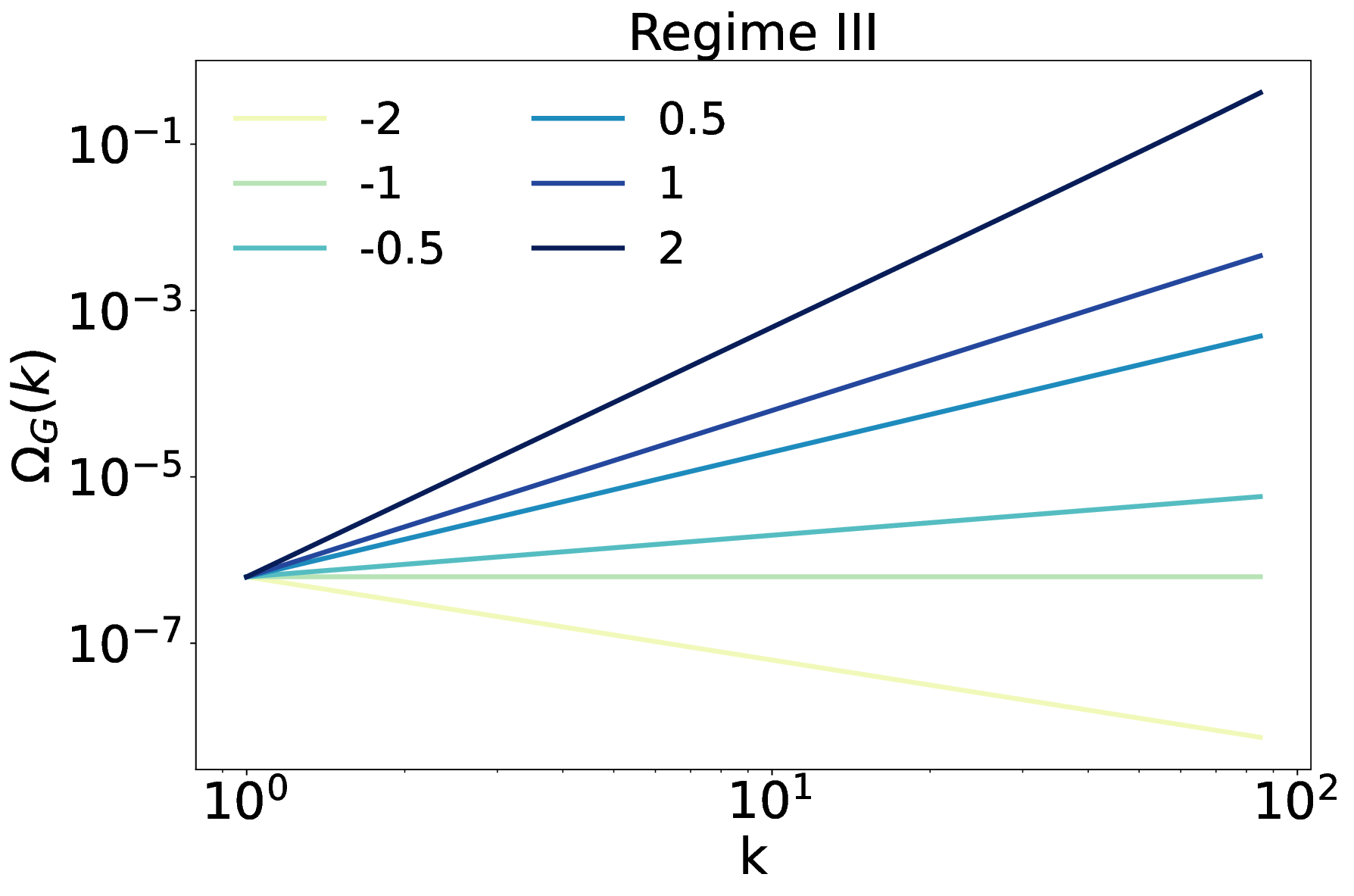}
		\caption{\label{fig:regime3}}
	\end{subfigure}
	\caption{\label{fig:symmetry} 
	a) Generalised energy spectrum \vd{$E_G(k)$} for $\beta = -10^2$ and $\gamma=10^7$ in Regime I. 
	b) Generalised enstrophy spectrum \vd{$\Omega_G(k)$ for $\beta=10^7$ and $\gamma=-10^2$ in Regime III}.}
\end{figure}

To demonstrate this mapping, we plot Eqs. \eqref{eq:equilspec} for different values of $\alpha$ and fixed values of $\beta$ and $\gamma$ in Fig. \ref{fig:symmetry}. 
\vd{We fix the 
wavenumbers $k_{\min} = 1$ and $k_{\max} = 85$.}
The behaviour in regime I for different $\alpha$ is shown in Figure \ref{fig:regimes1}, where we fix $\beta=-10^2$ and $\gamma=10^7$. While for regime III, the behaviour for different $\alpha$ is shown in Fig. \ref{fig:regime3}, where the values of $\beta=10^7$ and $\gamma=-10^2$ are exchanged compared to Figure \ref{fig:regimes1}. By comparing the figures, we can see the mapping from regime I to regime III as \vd{$\alpha \mapsto -\alpha$}, the behaviour of the spectra $E_G (k)$ and $\Omega_G (k)$ are identical. \vd{The existence of this symmetry implies} that it is sufficient to focus only on the positive \vd{$\alpha$} values.
Secondly, the exchange in behaviour \vd{between} $E_G (k)$ and $\Omega_G (k)$ as \vd{$\alpha \mapsto -\alpha$} has similarities to the behaviour seen in the forced-dissipative system, \vd{where 
the cascade directions of the conserved quantities reverse when \vd{$\alpha \mapsto -\alpha$}, \citep{jha2025cascades}.}

\subsection{\vd{The limit $\alpha \rightarrow 0$: multiple scale analysis}}
\label{sec:mult_scales}
Next we aim to understand the behaviour of the system as $\alpha \rightarrow 0$. Using the result from the previous section, we can restrict our study to positive values of $\alpha$. 
First let us consider the generalised vorticity equations \eqref{eq:gmodel} and \eqref{eq:alpha} for $\alpha = 0$,
where Eq. \eqref{eq:alpha} gives $q = \psi$, and Eq. \eqref{eq:gmodel} becomes
\begin{equation}
  \partial_t \psi + J(\psi, \psi) = 0.
\end{equation}
Since $ J(\psi, \psi) = 0 $, this implies $ \partial_t \psi = 0 $, i.e. the field is stationary and no energy transfer occurs. While at $\alpha = 0$ the dynamics 
\vbj{is} trivial, in the limit of $\alpha \rightarrow 0$ we expect the dynamics to slow down due to the weak nonlinear term \citep{jha2025cascades}. To find the limiting behaviour of the system, we derive the leading order evolution equation that governs the dynamics of the generalised vorticity using the method of multiple scales. 

We take $\alpha$ to be a small parameter, $0 < \alpha \ll 1$, which sets the slow nonlinear time scale. 
We then 
\vd{consider} two different time scales, 
the fast timescale $t$ and the slow timescale $\tau = \alpha t$. 
Physically, the fast timescale can be thought as the eddy turnover time $t \sim 1 / (U k_{\min})$, where $U$ is the rms velocity. 
The field $\wh \psi$ is then assumed to depend on both timescales as follows
\begin{equation}
\wh\psi = \wh\psi({\bf k}, t, \tau),
\end{equation}
so that the time derivative transforms to
\begin{equation}
\partial_t = \partial_{t} + \alpha \partial_\tau.
\end{equation}
Expanding $\wh\psi$ in powers of $\alpha$, we have
\begin{equation}
  \wh\psi = \wh\psi_0({\bf k}, t, \tau) + \alpha \wh\psi_1({\bf k}, t, \tau) + O(\alpha^2).
  \label{eq:psi_exp}
\end{equation}
Moreover, for $\alpha\ll 1$ Eq. \eqref{eq:qk} becomes
\begin{equation}
  \wh q = \wh\psi + \alpha \vd{\wh L} \wh\psi + O(\alpha^2),
  \label{eq:q_exp}
\end{equation}
where we use the expansion $k^{\alpha} = 1 + \alpha \vd{\wh L} + \mathcal{O}(\alpha^2)$, with \vd{$\wh L = \log{k}$}. If the system is truncated at a finite $k$ before taking \vd{the limit $\alpha \to 0$, this means that $\alpha \wh L \ll 1$ in the expansion \eqref{eq:q_exp}}.

Substituting the expansions \eqref{eq:psi_exp} and \eqref{eq:q_exp} into \eqref{eq:gmodel}, we collect terms order by order in $\alpha$.
At leading order $\mathcal{O}(1)$, \vd{in Fourier space we obtain}
\begin{equation}
  \vd{\partial_{t} \wh\psi_0 + J\wh{(\psi_0, \psi_0)} = 0},
\end{equation}
where $J\wh{(a, b)}$ is the Fourier transform of the Jacobian term $J (a, b)$.
Since \vd{$J\wh{(\psi_0, \psi_0)} = 0$}, we find that $ \partial_{t} \wh\psi_0 = 0 $ and so $\wh \psi_0$ is independent of the fast timescale $t$, though it can evolve on the slow timescale $\tau$. Thus, the leading-order solution 
is a quasi-stationary field evolving on the slow timescale
\begin{equation}
  \wh\psi_0 = \wh\psi_0({\bf k}, \tau).
\end{equation}
At 
\vd{leading} order the modes are decoupled and no energy exchange occurs between modes.
At the next order $\mathcal{O}(\alpha)$, we obtain
\begin{equation}
  \pd_{t} (\vd{\wh L \wh\psi_0} + \wh\psi_1) + \pd_\tau \wh\psi_0 + J\wh{(\psi_0, \vd{L\psi_0})} + J\wh{(\psi_0,\psi_1)} + J\wh{(\psi_1,\psi_0)} = 0,
\end{equation}
\vd{where $L$ is the \vbj{real space representation of}
the operator $\wh L = \log k$, which gives a nonlocal kernel in physical space.}
Since $ \pd_t \vd{\wh L \wh\psi_0} = \pd_{t} \wh\psi_0 =  0 $, this simplifies to:
\begin{equation}
\pd_{t} \wh\psi_1 + J\wh{(\psi_0,\psi_1)} + J(\wh{\psi_1,\psi_0)} = - \pd_\tau \wh\psi_0 - J\wh{(\psi_0, \vd{L\psi_0})}.
\end{equation}
This equation describes the evolution of the correction $\wh \psi_1$ on the fast timescale $t$ forced by the right-hand side (RHS), which
depends only on the slow timescale $\tau$.

The solvability condition should ensure that $\psi_1$ does not grow unbounded over time. Thus, we require the projection of the forcing $- \pd_\tau \wh\psi_0 - J\wh{(\psi_0, \vd{L\psi_0})}$ onto the neutral modes of the operator in the left-hand side to vanish.
The solvability condition is implemented by averaging over the fast timescale $t$, yielding the slow evolution equation
\begin{equation}
 \pd_\tau \wh\psi_0  + J\wh{(\psi_0, \vd{L\psi_0})} = 0. 
 \label{eq:reduced_model}
\end{equation}
This equation governs the slow evolution of $\wh \psi_0$ on the slow timescale $\tau$ and the term $J\wh{(\psi_0, \vd{L\psi_0})}$ represents the weak nonlinear interactions between modes due to the weak coupling introduced by $\alpha \ll 1$. These weak interactions cause a gradual redistribution of energy among the modes, which can eventually lead to thermalisation. 

Since the slow dynamics occur on the timescale $\tau = \alpha t$, the characteristic time for the nonlinear interactions of the generalised vorticity equations \eqref{eq:gmodel}, \eqref{eq:alpha} is thus given by 
\begin{equation}
  \tau_{_{NL}} \propto \frac{1}{\alpha},
  \label{eq:interaction_time}
\end{equation}
implying that as $\alpha \to 0$ the energy exchange amongst the modes becomes slower,  leading the system to reach statistical equilibrium at longer timescales.  

\vd{For systems with $k_{\max} \gg 1$, where essentially we take the thermodynamic limit prior to the $\alpha \rightarrow 0$ limit, the expansion \eqref{eq:q_exp} fails since higher order contributions become important. 
The above analysis is valid as long as $\alpha \wh L \ll 1$, which is true for the truncated equations with a finite $k_{\max} \ll 1/\alpha$. 
Thus, the reduced model \eqref{eq:reduced_model} and the nonlinear time scale \eqref{eq:interaction_time} are valid because we take first the $\alpha \rightarrow 0$ limit for the system truncated at a finite $k$.}
\subsubsection{\vd{Statistical equilibrium solutions of the reduced model}}
\vd{The reduced model \eqref{eq:reduced_model} can be numerically solved to find the behaviour of the full nonlinear model \eqref{eq:gmodel} in the limit $\alpha \rightarrow 0$. 
Equation \eqref{eq:reduced_model} has the following quadratic conserved quantities,
\begin{equation}
  E_R = \avg{\psi_0^2}, \quad \Omega_R = \langle \psi_0 L \psi_0 \rangle,
  \label{eq:invariants_R}
\end{equation}
where $E_R$ and $\Omega_R$ play the role of the generalised energy and enstrophy of the reduced model, respectively. 
Furthermore, we can derive the statistical equilibrium states corresponding to these conserved quantities by repeating the steps of the canonical ensemble theory discussed in section \ref{sec:theory}. 
In this case, we find the generalised energy and enstrophy spectra of the reduced model to be given by
\begin{align}
	  E_R (k) = \frac{2\pi k}{\beta_R + \gamma_R \log k}, \quad
\Omega_R(k) = \frac{2\pi k \log k}{\beta_R +\gamma_R \log k},
\label{eq:equilspecred}
\end{align}
where $\beta_R$ and $\gamma_R$ are the inverse temperature parameters of the canonical ensemble probability density, which is determined by the invariants \eqref{eq:invariants_R} similar to Eq. \eqref{eq:pdf}.}

\section{Numerical Results}
\label{sec:results}

\subsection{Numerical set-up}

To study the relaxation of the generalised vorticity equations towards equilibrium and the effect of $\alpha$ on the dynamics, we perform DNS in a periodic square domain of size $[0,2\pi] \times [0,2\pi]$ of the truncated equations. We numerically integrate Eqs. \eqref{eq:gmodel}, \eqref{eq:alpha} using the pseudospectral method \citep{GottliebOrszag1977} and a third-order Runge-Kutta scheme for the time advancement. The aliasing errors are removed with the 2/3 dealiasing rule, ensuring that the maximum wavenumber is $k_{\max}=N/3$ \citep{mpicode2005}, where $N$ denotes the number of aliased modes in the $x$ and $y$ directions. We fix the time-step to $dt=10^{-4}$ and we explore numerical resolutions $N^2 = 256^2$ and $512^2$. The flow is always initialised with a generalised energy spectrum $E_G(k) \propto k^{-2}$ for $k \in [1,4]$ and zero elsewhere. 
The most interesting statistical equilibrium solutions are the negative temperature states. 
\vd{For this reason, we focus} our study to the solutions which are in regime I. With that in mind, at $t=0$ we 
\vd{fix the amplitude 
of} the generalised energy $E_G(t=0) = 1$ and \vd{$\Omega_G(t=0)$ is 
fixed by the energy} spectrum of the initial conditions and \vd{the value of $\alpha$. For the values of $\alpha$ considered in this study, we find that $k_c = (\Omega_G / E_G)^{1/\alpha} \in [1.0001,1.67]$.}  
The values of $\beta$ and $\gamma$ for each $\alpha$ are obtained by numerically solving for $E_G(t=0)=\sum_{k}E_G(k)$ and $\Omega_G(t=0)=\sum_{k}\Omega_G(k)$. 
Tests were conducted for other initial conditions as well, and they are discussed in Appendix \ref{app:B}.

\subsection{Spectral dynamics}

\begin{figure}
	\centering
	\begin{subfigure}[h]{0.45\textwidth}
		\includegraphics[width=\linewidth]{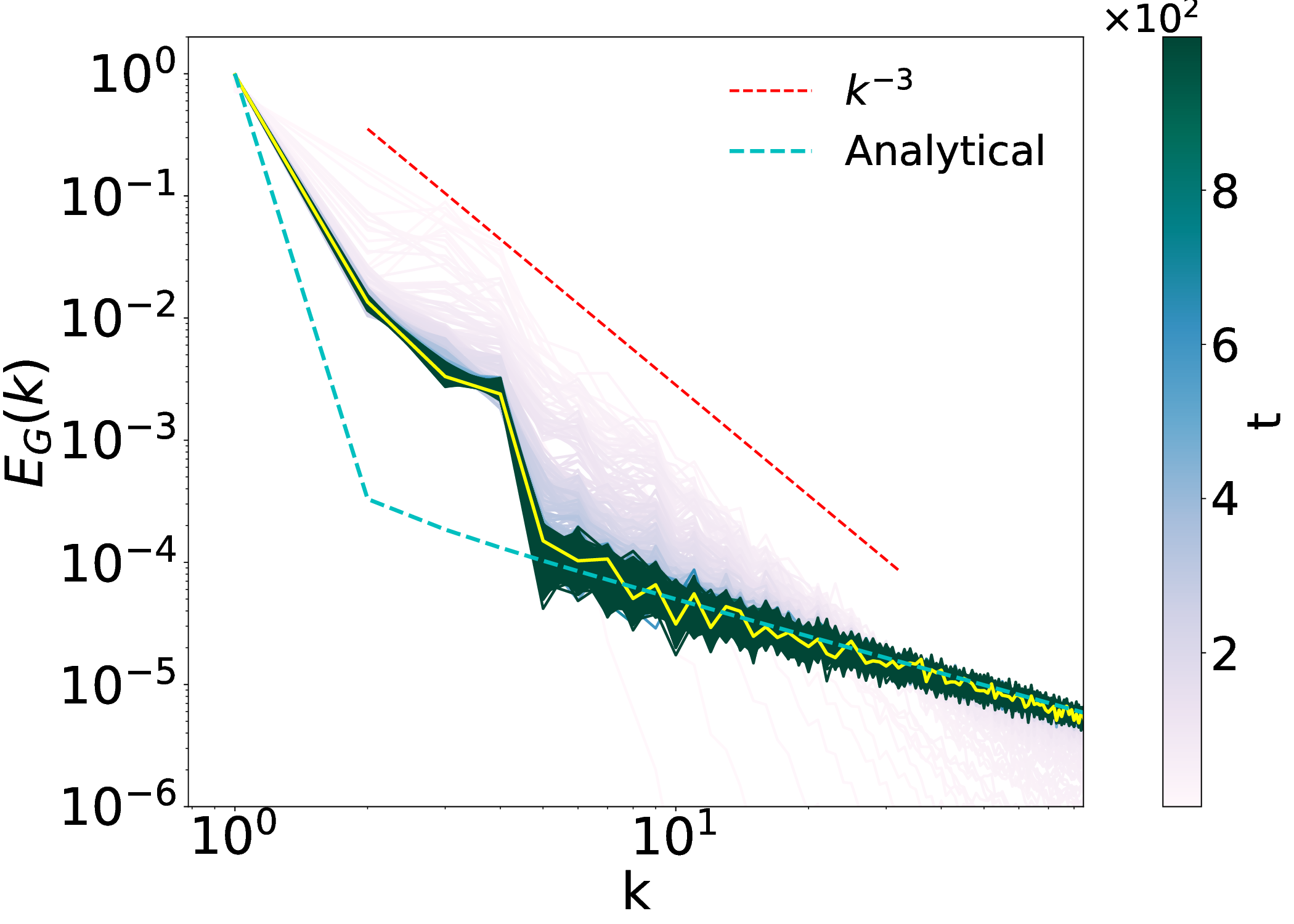}
		\caption{\label{fig:spec2}}
	\end{subfigure}
	\begin{subfigure}[h]{0.45\textwidth}
		\includegraphics[width=\linewidth]{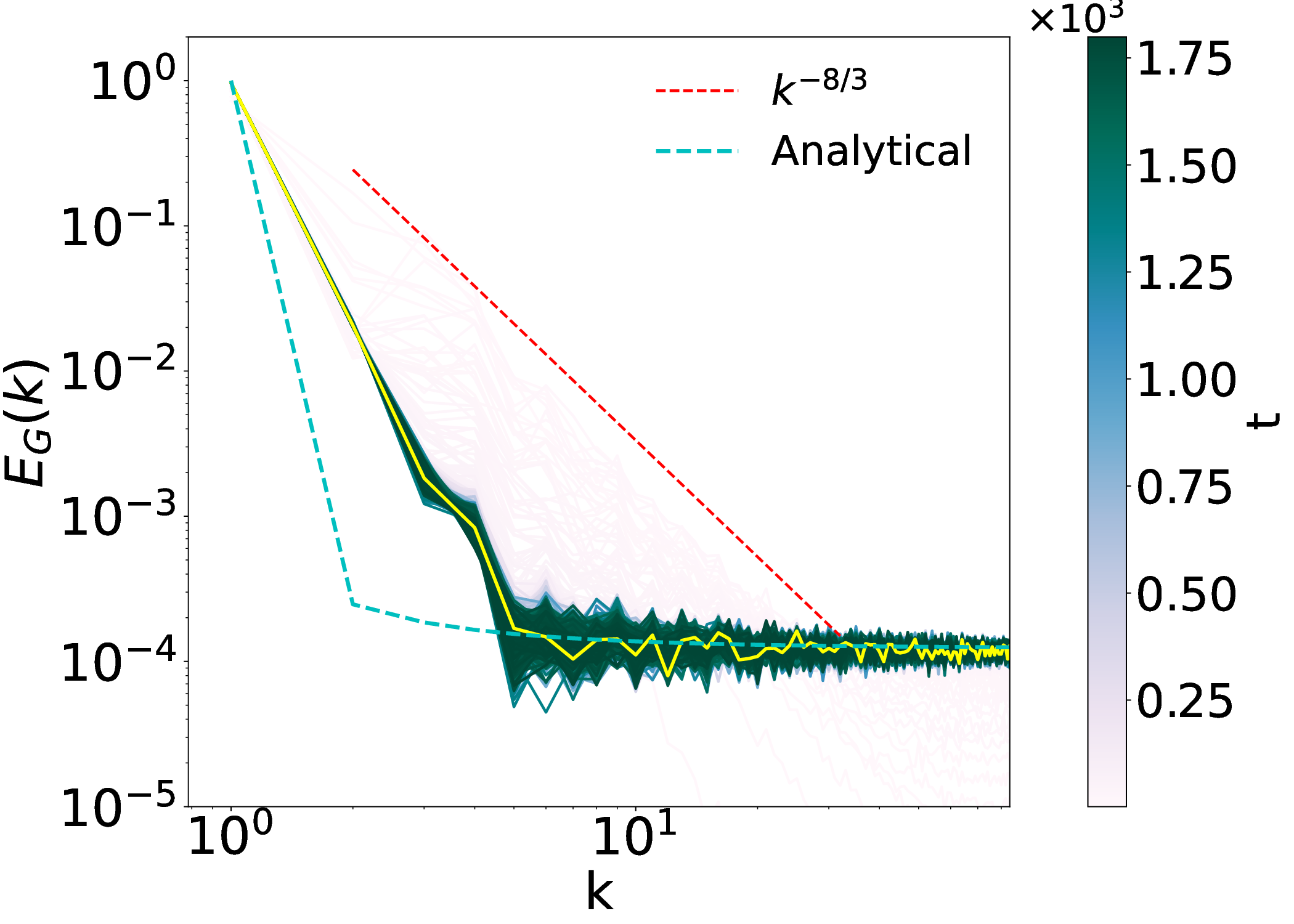}
		\caption{\label{fig:spec1}}
	\end{subfigure}
	\begin{subfigure}[h]{0.45\textwidth}
		\includegraphics[width=\linewidth]{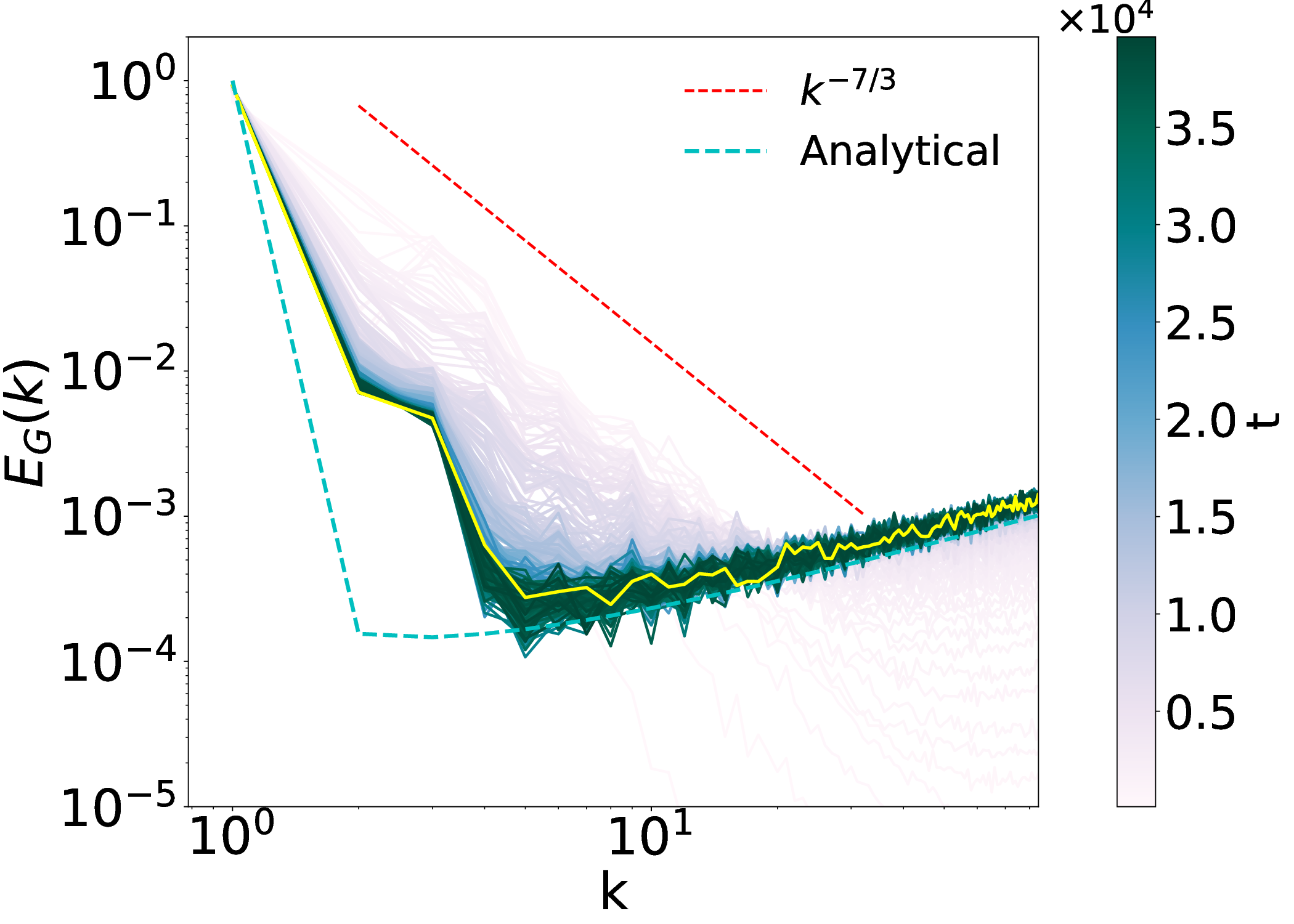}
		\caption{\label{fig:spec03}}
	\end{subfigure}
	\begin{subfigure}[h]{0.45\textwidth}
		\includegraphics[width=\linewidth]{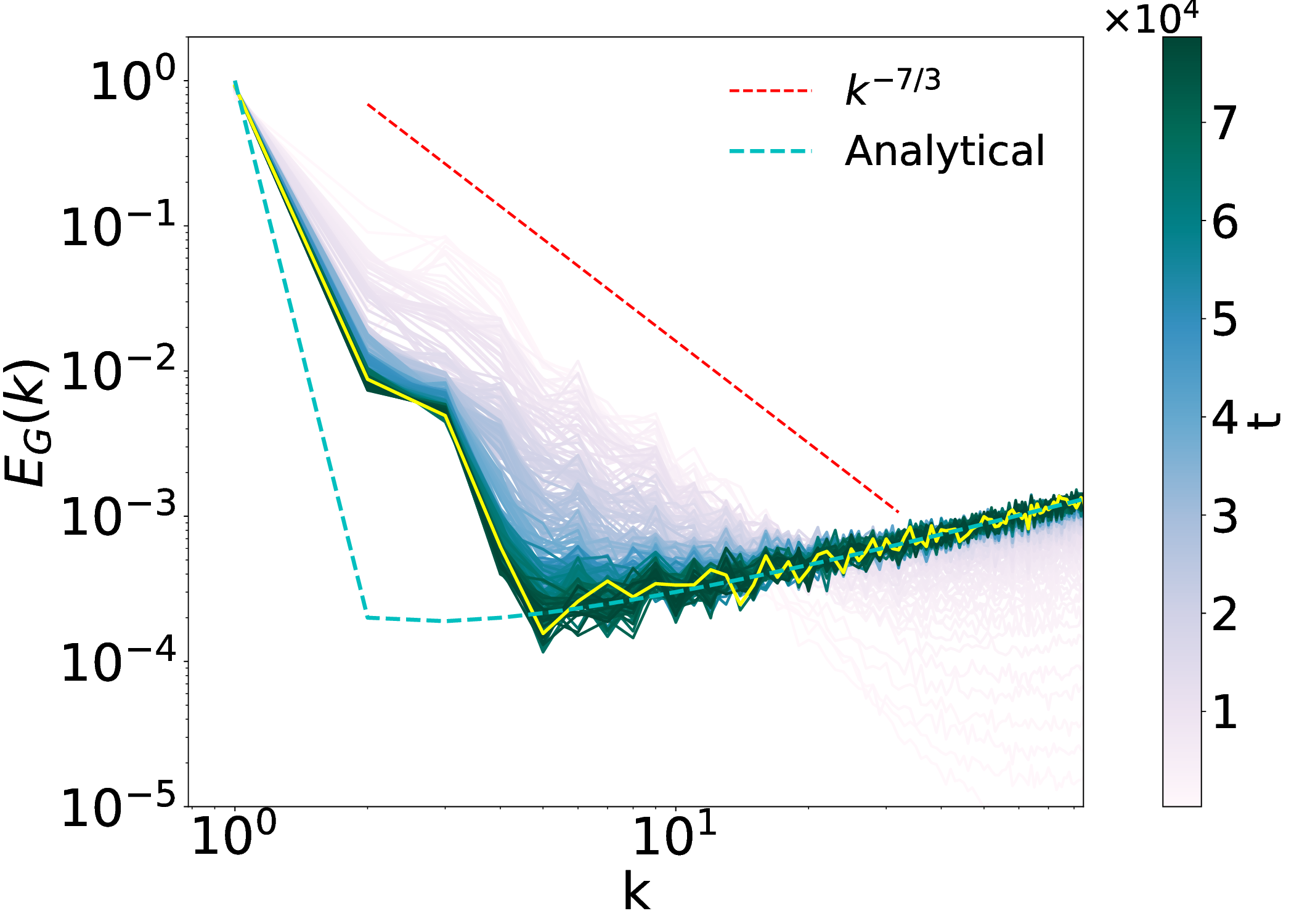}
		\caption{\label{fig:spec001}}
	\end{subfigure}
	\caption{\label{fig:spectimes} 
	 Time evolution of the generalised energy spectra for different values of $\alpha$. a) $\alpha=2.0$, b) $\alpha=1.0$ c) $\alpha=0.003$, d) $\alpha=0.001$.  
	\vd{Transient spectra follow the power-law $k^{-(7+\alpha)/3}$ (red dashed lines) at intermediate scales before the system relaxes toward equilibrium. 
	The sequence demonstrates the progressive spread of thermalised modes from small to large scales. 
	Cyan dashed lines show the statistical equilibrium spectra \eqref{eq:equilspec}}.}
\end{figure}

Starting from the narrow-band initial condition concentrated 
\vd{in wavenumbers $k \in [1,4]$}, the system evolves to a partially equilibrated solution for all values of $\alpha$. Fig.~\ref{fig:spectimes} shows \vd{the generalised energy spectra of the} intermediate states 
as time evolves \vd{following the colourbar and the latest time spectrum} in yellow for different values of $\alpha$.
At intermediate times, a power-law spectrum $E_G(k) \propto k^{-(\alpha + 7)/3}$ emerges, which is indicated by the red dashed lines in Fig. \ref{fig:spectimes}. This scaling is consistent with the predictions for the dissipative generalised vorticity equations following \cite{K41} scaling arguments \citep{Smithetal2002, jha2025cascades}.

This Kolmogorov-like transient behaviour has also been identified in truncated 3D Euler and inviscid SQG \citep{cichowlas2005effective,krstulovic2008two,krstulovic2009cascades,TeitelbaumMininni2012}. During the \vd{relaxation} to equilibrium a viscous-like inertial range develops through which thermalisation proceeds to lower wavenumbers. As time advances, thermalisation first develops at small scales and then progressively spreads towards larger scales, until the spectrum approaches the partial equilibrium state. 

The analytical equilibrium spectra \eqref{eq:equilspec} derived in section \ref{sec:theory} are plotted in Fig.~\ref{fig:spectimes} in dashed cyan color. As the large wavenumber modes thermalise, the numerical spectra start approaching the analytical spectra.
Fig.~\ref{fig:spectimes} shows that at late times the large wavenumbers of the computed spectra are well predicted by the canonical ensemble theory \vd{derived in section \ref{sec:theory}}, but the deviation is clearly noticeable in the small wavenumbers. This deviation, which is due to the partial thermalisation, has also been observed in previous studies \citep{FoxOrszag1973,venaille2015violent} when $k_c/k_{\max}$ is small \citep{Agouaetal2025}. \vd{This} is attributed to the fact that the canonical ensemble theory 

\vd{is not able to capture} solutions in the condensate regime, especially \vd{the dynamics of the} modes close to the smallest wavenumber of the domain. \vd{Nevertheless} the behaviour at larger wavenumbers \vd{$k \gtrsim 5$} follow the analytical spectrum well for all values of $\alpha$ examined in this study.
\subsection{\vd{Thermalisation}}
\label{sec:tau_th}

To quantify the time taken to reach a partial thermalised state, we define a quantity $E_{th}$ using the sum of the generalised energy $E_G$ contained in 
wavenumber modes $k \geq k_{th}$, \vd{namely}
\begin{equation}
    \vd{E_{th}(t)} = \sum_{k = k_{th}}^{k_{\max}} E_G(k,t).
    \label{eq:Eth}
\end{equation}
Here $k_{th}$ is a wavenumber, which is used to denote the modes above which thermalisation has taken place in the system, $k \geq k_{th}$,  \citep{TeitelbaumMininni2012}. At long times we expect $E_{th}$ to reach a statistical steady state, whose \vd{average value will depend on $\alpha$. To remove this dependence, we subtract from the signal its long-time average, computed over the time interval $T$ defined by $0.9 t_{\max} \leq t \leq t_{\max}$, where $t_{\max}$ denotes the maximum integration time. We denote the temporal average of $E_{th}$ as $\avg{E_{th}}_T$.
The time $t_{\max}$ is chosen sufficiently large so that the average $\avg{E_{th}}_T$, taken over an interval spanning several tens of the turnover time $L/U$, is statistically converged.} 

\begin{figure}
	\centering
	\begin{subfigure}[h]{0.49\textwidth}		
		\includegraphics[width=\linewidth]{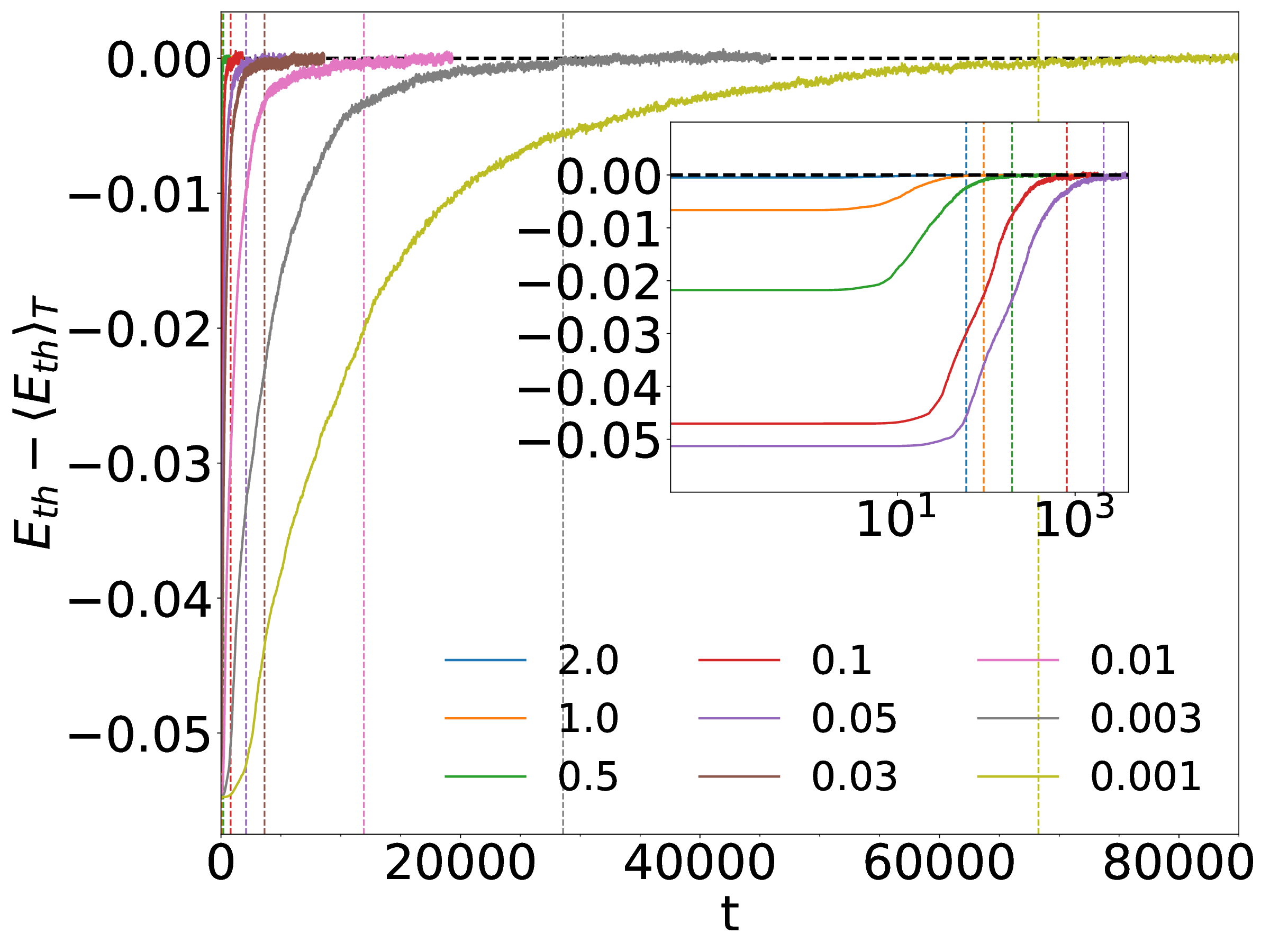}
		\caption{\label{fig:Eth}}
	\end{subfigure}
	\begin{subfigure}[h]{0.49\textwidth}		
		\includegraphics[clip,width=\linewidth]{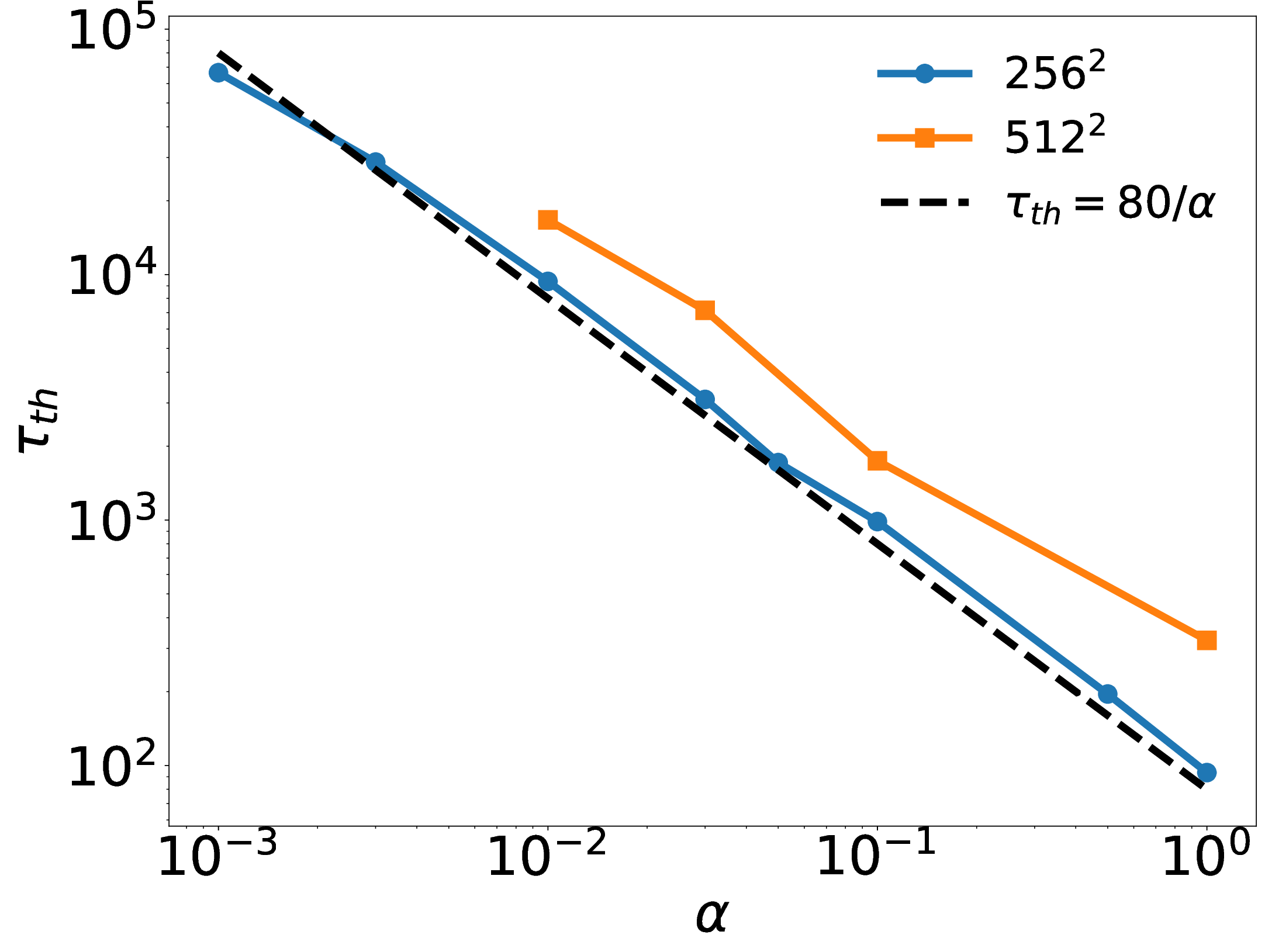}
		\caption{\label{fig:tau_th}}
	\end{subfigure}
	\caption{\label{fig:ts} 
	\vd{a) Time series of adjusted thermalised energy $E_{th}(t) - \avg{E_{th}(t)}_T$ for the wavenumber modes $k \geq k_{th}$, with $k_{th} = 30$ and for different values of $\alpha$.} The dashed vertical lines indicate the onset of thermalisation according to the threshold criterion. 
	b) Thermalisation time $\tau_{th}$ as a function of $\alpha$, shown for simulations with $256^2$ and $512^2$ \vd{resolution. The black dotted line indicates the power-law $\tau_{th} = 80/\alpha$ which is obtained using DNS of the reduced model.}
}
\end{figure}
 
Figure \ref{fig:Eth} shows the adjusted thermalised energy $E_{th}(t) - \avg{E_{th}(t)}_T$ for different values of $\alpha$ and for $k_{th} = 30$. 
At early times, the initial conditions do not excite modes beyond $k_{th}$ implying that $E_{th}$ starts from zero. Hence, the quantity $E_{th}(t) - \avg{E_{th}(t)}_T$ is negative initially and grows monotonically as nonlinear interactions redistribute energy from the initially excited large-scale band toward higher wavenumbers. 
 
As time evolves, the modes start to thermalise gradually and the \vd{adjusted thermalised energy} $E_{th}(t) - \avg{E_{th}(t)}_T$ starts to saturate \vd{toward zero} as shown in Fig. \ref{fig:Eth}. The inset shows the time series for values \vd{$0.05 \leq \alpha \leq 2$} which saturate fast, while those which have \vd{$\alpha < 0.05$} take a longer time to saturate. 
\vd{This behaviour arises because the nonlinear term becomes progressively weaker as $\alpha$ is reduced, as discussed in Section \ref{sec:mult_scales}.} To quantify the onset of partial thermalisation, we find the time \vd{$t = \tau_{th}$} at which $E_{th}(t)$ starts saturating and becomes statistically stationary. 

\vd{Practically, we determine the partial thermalisation time $\tau_{th}$ as the earliest time at which the condition $| E_{th}(t) - \avg{E_{th}(t)}_T | \leq 10^{-6}$ is satisfied. We have verified that the value $10^{-6}$ of the threshold does not affect 
our results.} The values of $\tau_{th}$ are shown by vertical dashed lines in Fig. \ref{fig:Eth} for each value of $\alpha$. The \vd{same} analysis is also carried out for 
\vd{the reduced model \eqref{eq:reduced_model} 
to capture the behaviour of $\tau_{th}$ 
in the limit of $\alpha \rightarrow 0$.}

Due to inherent fluctuations in the system, an accurate way to define the partial thermalisation time $\tau_{th}$ \vd{would be through an ensemble average over multiple realisations.
Owing to the high computational cost involved, we present results from a single ensemble of simulations.} 

\vd{Now, we aim to understand the dependence of $\tau_{th}$ on $\alpha$, by plotting $\tau_{th}$ as a function of $\alpha$ in logarithmic scale in Fig. \ref{fig:tau_th}.} 
To compare with the reduced model \eqref{eq:reduced_model}, we divide the $\tau_{th}$ obtained from the reduced model by $\alpha$, shown as dashed line \vd{in the plot. Figure \ref{fig:tau_th}} reveals a clear $\tau_{th} \propto 1/\alpha$ \vd{power-law} scaling, indicating that $\tau_{th}$ diverges in the limit $\alpha \to 0$ and that the reduced model captures correctly this behaviour along with the expected prefactor. 
This validates the slow dynamics of the multiple scale analysis in Section \ref{sec:mult_scales}, implying that the relevant time scale at which energy transfer occurs at small $\alpha$, scales like \vd{$1/\alpha$}. The same scaling persists at a higher resolution of $512^2$ (see Fig. \ref{fig:tau_th}) \vd{for $\alpha \leq 0.1$}, where the wave number $k_{th} = 30$ is kept the same, suggesting resolution-independent behaviour. 
\vd{The $1/\alpha$ scaling is expected to hold only for sufficiently small values of $\alpha$, and deviations naturally arise as $\alpha$ becomes order one.}
The partial thermalisation time increases with resolution for fixed $\alpha$ (see Fig. \ref{fig:tau_th}), because a larger number of modes is required to equilibrate. This also makes the investigation of very small values of $\alpha$ computationally \vd{challenging, due to the rapid growth of $\tau_{th}$ at high resolution.}

Finally, we have also considered different initial conditions to \vd{test} if they affect the $\tau_{th} \propto 1/\alpha$ scaling. In Appendix \ref{app:B}, we show that the scaling does not depend on the initial conditions or on the specific choice of $k_{th}$, as long as a clear 
separation exists between the initially driven modes and the wavenumbers at which thermalisation is assessed. 
\vd{The presence of a clear 
separation allows us to determine $\tau_{th}$ unambiguously by sharply identifying the onset of partial thermalisation in the system.}

\section{Conclusions}
\label{sec:conc}

In this article, we focus on the 
\vd{canonical ensemble theory and the relaxation toward statistical equilibrium of generalised 2D flows.} 
The parameter $\alpha$, which relates the generalised vorticity with the steamfunction \vd{controls} the strength of the nonlinearity. 
We find that the \vd{statistical} equilibrium solutions 
of the truncated system show $\alpha \mapsto - \alpha$ symmetry, where changing the sign of $\alpha$ is equivalent of \vd{exchanging $E_G \leftrightarrow \Omega_G$}. In the dissipative system, changing the sign of $\alpha$ leads to a change in the direction of the cascade of the quantities \vd{$E_G$ and $\Omega_G$ \citep{jha2025cascades}}. Hence, both the inviscid and the dissipative system show some aspects of the \vd{$\alpha \mapsto -\alpha$ symmetry.}

By numerically integrating the truncated equations, we can study the time taken to reach a partially thermalised state. The 
\vd{relaxation to equilibrium} proceeds via development of viscous like transient states, \vd{while} thermalisation proceeds to lower wavenumbers by the emergence of an effective viscosity. 
Defining $E_{th}$ as the energy content in the wavenumbers beyond a cutoff wavenumber $k_{th}$, we quantify the partial thermalisation time $\tau_{th}$ using \vd{the statistical stationarity of $E_{th}$ time series}. We observe a slowdown of the 
\vd{relaxation to equilibrium}, with $\tau_{th}$ diverging as $1/\alpha$ \vd{in the limit} $\alpha \to 0$. 
\vd{This behaviour is well captured by a reduced model in which the generalised vorticity-streamfunction relation is logarithmic.}

Although we understand reasonably well the scaling $\tau_{th} \propto 1/\alpha$, obtained for a specific set of initial conditions, we have not explored the full space of possible initial conditions. 
In this study, the simulations are initialised in regime~I, where condensate formation occurs, leading the system into long-lived quasi-equilibrium states that remain far from the statistical equilibrium predicted from the canonical ensemble theory, as illustrated in Fig. \ref{fig:spectimes} and reported in previous works \citep{venaille2015violent,Agouaetal2025}. 
This behaviour persists even as nonlinear interactions become progressively weaker in the limit of $\alpha \to 0$. By contrast, the initial conditions we tested in regimes~II and~III, but not reported here, relax toward statistical equilibrium states that are close to the canonical ensemble predictions. 
\vd{Both the reduced model and the numerical simulations are performed in the weakly nonlinear regime and at finite system size, under the condition $\alpha k_{\max} \ll 1$. 
A natural extension of the present study is to explore the thermodynamic limit $k_{\max} \gg 1$ as $\alpha \rightarrow 0$ with either $\alpha k_{\max} \to \text{constant}$ or $\alpha k_{\max} \to \infty$. Although the strength of individual nonlinear interactions vanishes in the limit $\alpha \to 0$, the number of interacting modes grows as $k_{\max}^2$, 
potentially compensating for the weakening of triadic interactions and giving rise to collective effects absent in finite-size systems. Such an interplay may lead to deviations from the $\tau_{th} \propto 1/\alpha$ scaling observed here and motivates the development of theoretical frameworks capable of addressing this regime. More broadly, while we have highlighted similarities between the dissipative and the inviscid system under the transformation $\alpha \rightarrow -\alpha$, it would be of considerable interest to investigate whether analogous symmetry principles arise in other systems that exhibit transitions in the direction of cascades of conserved quantities.}
\section{Acknowledgements}
Simulations used computational resources of SAGAR High Performance Cluster at Space Applications Centre(SAC), Indian Space Research Organisation (ISRO), Ahmedabad.
\section{Declaration of Interests}
The authors report no conflict of interest.
\appendix

\section{\vd{Symmetry of conserved quantities}} 
\label{appA}
The mean generalised energy \vd{$E_G$ and enstrophy $\Omega_G$ can be written as follows
\begin{equation}
    E_G = \int_{k_{min}}^{k_{max}}\frac{2\pi k}{\beta+\gamma k^{\alpha}}dk, \quad
    \Omega_G
     = \int_{k_{min}}^{k_{max}}\frac{2\pi k^{\alpha+1}}{\beta+\gamma k^{\alpha}} dk
  \label{eq:meanEO_G}
\end{equation}
}

We show the expression of these integrals in Table \ref{tbl:E_Omega_alpha} for $\alpha=2,1,-1$ and $-2$,  \vd{to elucidate the symmetry 
$\alpha \rightarrow -\alpha$ mathematically}.
As seen from the table, the quantities $E_G$ and $\Omega_G$ are exchanged along with the inverse temperatures $\beta$ and $\gamma$ as $\alpha \rightarrow -\alpha$.
\begin{table*}
\centering
\begin{tabular}{ccc}
\hline
\hline
$\alpha$ & $E_G$ & $\Omega_G$ \\
\hline
$2$ 
&
$\displaystyle 
\frac{\pi}{\gamma}
\left.
\log(\beta+\gamma k^2)
\right|_{k_{\min}}^{k_{\max}}
$
&
$\displaystyle 
\frac{\pi}{\gamma^2}
\left.
\bigl(\gamma k^2 - \beta \log(\beta+\gamma k^2)\bigr)
\right|_{k_{\min}}^{k_{\max}}
$
\\[10pt]
$1$ 
&
$\displaystyle 
\frac{2\pi}{\gamma^2}
\left.
\bigl(\gamma k - \beta \log(\beta+\gamma k)\bigr)
\right|_{k_{\min}}^{k_{\max}}
$
&
$\displaystyle 
\frac{\pi}{\gamma^3}
\left.
\bigl(2\beta^2 \log(\beta+\gamma k)
+ \gamma k(\gamma k - 2\beta)\bigr)
\right|_{k_{\min}}^{k_{\max}}
$
\\[10pt]
$-1$ 
&
$\displaystyle 
\frac{\pi}{\beta^3}
\left.
\bigl(2\gamma^2 \log(\beta k + \gamma)
+ \beta k(\beta k - 2\gamma)\bigr)
\right|_{k_{\min}}^{k_{\max}}
$
&
$\displaystyle 
\frac{2\pi}{\beta^2}
\left.
\bigl(\beta k - \gamma \log(\beta k + \gamma)\bigr)
\right|_{k_{\min}}^{k_{\max}}
$
\\[10pt]
$-2$ 
&
$\displaystyle 
\frac{\pi}{\beta^2}
\left.
\bigl(\beta k^2 - \gamma \log(\beta k^2 + \gamma)\bigr)
\right|_{k_{\min}}^{k_{\max}}
$
&
$\displaystyle 
\frac{\pi}{\beta}
\left.
\log(\beta k^2 + \gamma)
\right|_{k_{\min}}^{k_{\max}}
$
\\
\hline
\hline
\end{tabular}
\caption{\label{tbl:E_Omega_alpha}
\vd{Analytical expressions for the mean generalised energy $E_G$ and enstrophy $\Omega_G$ for selected values of the parameter $\alpha$. Note the manifestation of the $\alpha \mapsto -\alpha$ symmetry on the expressions, where $\beta \leftrightarrow \gamma$ and $E_G \leftrightarrow \Omega_G$.}}
\end{table*}

\section{\vd{Independence of $\tau_{th} \propto 1/\alpha$ on initial conditions and $k_{th}$}}
\label{app:B}

\begin{figure}
	\centering
	\includegraphics[width=0.5\linewidth]{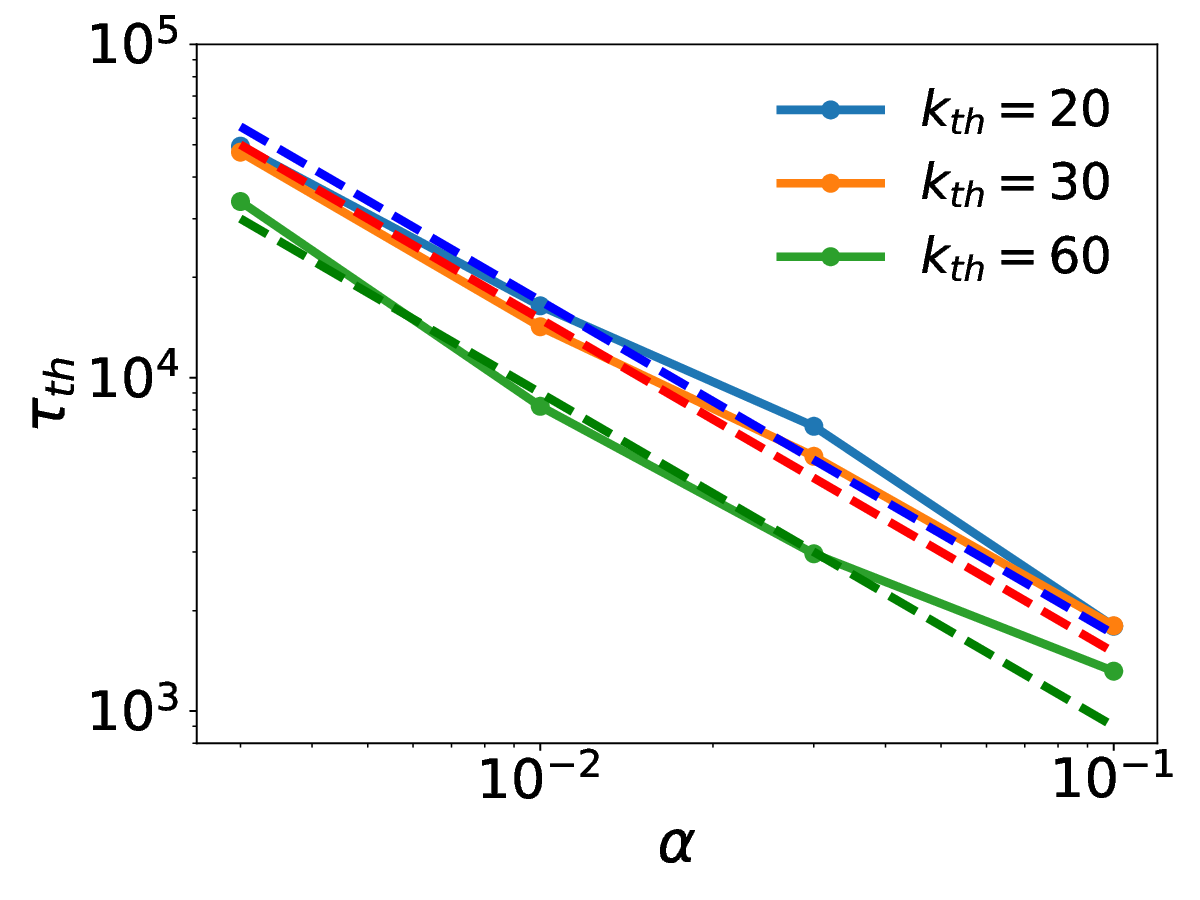}
    \caption{\vd{Partial} thermalisation time $\tau_{th}$ as a function of $\alpha$ for initially excited modes $k \in [11,14]$ following a $E_G(k) \propto k^{-2}$ power-law spectrum. The  dashed reference lines indicate the scaling obtained from the reduced model for different values of $k_{th}$.
	\label{fig:ICs}
}
\end{figure}

Here we consider the effect of the initial conditions on the partial thermalisation time \vd{$\tau_{th}$} by initially exciting a different set of wavenumber modes, $k \in [11,14]$, which follow a similar scaling $E_G(k) \propto k^{-2}$. We perform DNS with resolution $256^2$ and we repeat the analysis described in section \ref{sec:tau_th} for $k_{th} = 20, 30$ and $60$ to test if there is any dependence on the thermalisation wavenumber that is assessed. 
In Fig. \ref{fig:ICs} we plot $\tau_{th}$ as a function of $\alpha$ in logarithmic scale to demonstrate that the $1/\alpha$ power-law scaling persists for this initial condition too and for \vd{other} values of $k_{th}$. 
This robustness indicates that the $\tau_{th} \propto 1/\alpha$ scaling is a genuine feature of the dynamics. \vd{The prediction from the reduced model Eq. \eqref{eq:reduced_model}, shown by dashed lines, agrees with the $\tau_{th}$ of the full system \eqref{eq:gmodel}-\eqref{eq:alpha}.} Other initial conditions not belonging in regime I have also been tested and they 
\vd{exhibit a $\tau_{th} \propto 1/\alpha$ scaling, not shown here.}
\vd{If $k_{th}$ is between the initially excited modes}, the fluctuations in the time series of $E_{th}$ do not allow us to clearly determine the value of the partial thermalisation time $\tau_{th}$.
For this reason we make sure that the \vd{initially excited modes are well separated from the $k_{th}$ that is assessed} in order to measure $\tau_{th}$ without any ambiguity. 
\bibliographystyle{jfm}
\bibliography{citations}

@article{venaille2015violent,
  title={Violent relaxation in two-dimensional flows with varying interaction range},
  author={Venaille, A and Dauxois, Thierry and Ruffo, Stefano},
  journal={Physical Review E},
  volume={92},
  number={1},
  pages={011001},
  year={2015},
  publisher={APS}
}

@article{weichman2022statistical,
  title={Statistical equilibrium principles in 2D fluid flow: from geophysical fluids to the solar tachocline},
  author={Weichman, Peter B and Marston, John Bradley},
  journal={Entropy},
  volume={24},
  number={10},
  pages={1389},
  year={2022},
  publisher={MDPI}
}

@article{onorato2023wave,
  title={Wave turbulence and thermalization in one-dimensional chains},
  author={Onorato, Miguel and Lvov, Yuri V and Dematteis, Giovanni and Chibbaro, Sergio},
  journal={Physics Reports},
  volume={1040},
  pages={1--36},
  year={2023},
  publisher={Elsevier}
}

@article{alexakis2019thermal,
  title={On the thermal equilibrium state of large-scale flows},
  author={Alexakis, Alexandros and Brachet, Marc-Etienne},
  journal={Journal of Fluid Mechanics},
  volume={872},
  pages={594--625},
  year={2019},
  publisher={Cambridge University Press}
}

@article{kraichnan1975statistical,
  title={Statistical dynamics of two-dimensional flow},
  author={Kraichnan, Robert H},
  journal={Journal of Fluid Mechanics},
  volume={67},
  number={1},
  pages={155--175},
  year={1975},
  publisher={Cambridge University Press}
}

@article{krstulovic2008two,
  title={Two-fluid model of the truncated Euler equations},
  author={Krstulovic, Giorgio and Brachet, Marc-{\'E}tienne},
  journal={Physica D: Nonlinear Phenomena},
  volume={237},
  number={14-17},
  pages={2015--2019},
  year={2008},
  publisher={Elsevier}
}

@book{salmon1998lectures,
  title={Lectures on geophysical fluid dynamics},
  author={Salmon, Rick},
  year={1998},
  publisher={Oxford University Press}
}

@book{majda2006nonlinear,
  title={Nonlinear dynamics and statistical theories for basic geophysical flows},
  author={Majda, Andrew and Wang, Xiaoming},
  year={2006},
  publisher={Cambridge University Press}
}

@article{Tabeling2002,
  title={Two-dimensional turbulence: a physicist approach},
  author={Tabeling, Patrick},
  journal={Physics reports},
  volume={362},
  number={1},
  pages={1--62},
  year={2002},
}

@article{BoffettaEcke2012,
  title={Two-dimensional turbulence},
  author={Boffetta, Guido and Ecke, Robert E},
  journal={Annual review of fluid mechanics},
  volume={44},
  pages={427--451},
  year={2012},
}

@article{Lapeyre2017,
   title = {Surface Quasi-Geostrophy},
   author = {Guillaume Lapeyre},
   journal = {Fluids},
   month = {3},
   volume = {2},
   year = {2017}
}

@article{bos2006dynamics,
  title={Dynamics of spectrally truncated inviscid turbulence},
  author={Bos, Wouter JT and Bertoglio, J-P},
  journal={Physics of Fluids},
  volume={18},
  number={7},
  year={2006},
  publisher={AIP Publishing}
}

@article{bourouiba2008model,
  title={Model of a truncated fast rotating flow at infinite Reynolds number},
  author={Bourouiba, L},
  journal={Physics of Fluids},
  volume={20},
  number={7},
  year={2008},
  publisher={AIP Publishing}
}

@article{TeitelbaumMininni2012,
   title = {Thermalization and free decay in surface quasigeostrophic flows},
   author = {Tomas Teitelbaum and Pablo D. Mininni},
   issue = {1},
   journal = {Physical Review E - Statistical, Nonlinear, and Soft Matter Physics},
   month = {7},
   volume = {86},
   year = {2012}
}

@article{LarichevMcWilliams1991,
   title = {Weakly decaying turbulence in an equivalent-barotropic fluid},
   author = {Vitaly D. Larichev and James C. McWilliams},
   journal = {Physics of Fluids A},
   pages = {938-950},
   volume = {3},
   year = {1991},
}

@article{Smithetal2002,
   author = {K. S. Smith and G. Boccaletti and C. C. Henning and I. Marinov and C. Y. Tam and I. M. Held and G. K. Vallis},
   journal = {Journal of Fluid Mechanics},
   month = {10},
   pages = {13-48},
   publisher = {Cambridge University Press},
   title = {Turbulent diffusion in the geostrophic inverse cascade},
   volume = {469},
   year = {2002}
}

@article{K41,
  author={Kolmogorov, Andrey Nikolaevich},
  title={The local structure of turbulence in incompressible viscous fluid for very large Reynolds},
  journal={Numbers. In Dokl. Akad. Nauk SSSR},
  volume={30},
  pages={301},
  year={1941}
}

@book{GottliebOrszag1977,
  title={Numerical analysis of spectral methods: theory and applications},
  author={Gottlieb, David and Orszag, Steven A},
  year={1977},
  publisher={SIAM}
}

@article{mpicode2005,
  author={G\'omez, D. O. and Mininni, P. D. and Dmitruk, P.},
  title={Parallel Simulations in Turbulent {MHD}},
  journal={Physica Scripta},
  volume={T116},
  pages = {123-127},
  year={2005}
}

@article{pierrehumbert1994spectra,
  title={Spectra of local and nonlocal two-dimensional turbulence},
  author={Pierrehumbert, Raymond T and Held, Isaac M and Swanson, Kyle L},
  journal={Chaos, Solitons \& Fractals},
  volume={4},
  number={6},
  pages={1111--1116},
  year={1994},
  publisher={Elsevier}
}

@article{held1995surface,
  title={Surface quasi-geostrophic dynamics},
  author={Held, Isaac M and Pierrehumbert, Raymond T and Garner, Stephen T and Swanson, Kyle L},
  journal={Journal of Fluid Mechanics},
  volume={282},
  pages={1--20},
  year={1995},
  publisher={Cambridge University Press}
}

@article{byrne2013height,
  title={Height-dependent transition from 3-D to 2-D turbulence in the hurricane boundary layer},
  author={Byrne, David and Zhang, Jun A},
  journal={Geophysical research letters},
  volume={40},
  number={7},
  pages={1439--1442},
  year={2013},
  publisher={Wiley Online Library}
}

@article{king2015upscale,
  title={Upscale and downscale energy transfer over the tropical P acific revealed by scatterometer winds},
  author={King, Gregory P and Vogelzang, Jur and Stoffelen, Ad},
  journal={Journal of Geophysical Research: Oceans},
  volume={120},
  number={1},
  pages={346--361},
  year={2015},
  publisher={Wiley Online Library}
}

@article{shao2023physical,
  title={A physical model for the observed inverse energy cascade in typhoon boundary layers},
  author={Shao, Xin and Zhang, Ning and Tang, Jie},
  journal={Geophysical Research Letters},
  volume={50},
  number={18},
  pages={e2023GL105546},
  year={2023},
  publisher={Wiley Online Library}
}

@article{young2017forward,
  title={Forward and inverse kinetic energy cascades in Jupiter’s turbulent weather layer},
  author={Young, Roland MB and Read, Peter L},
  journal={Nature Physics},
  volume={13},
  number={11},
  pages={1135--1140},
  year={2017},
  publisher={Nature Publishing Group UK London}
}

@article{Kraichnan1967,
  title={Inertial ranges in two-dimensional turbulence},
  author={Kraichnan, Robert H},
  journal={The Physics of Fluids},
  volume={10},
  number={7},
  pages={1417--1423},
  year={1967}
}

@article{ScottWang2005,
  title={Direct evidence of an oceanic inverse kinetic energy cascade from satellite altimetry},
  author={Scott, Robert B and Wang, Faming},
  journal={Journal of Physical Oceanography},
  volume={35},
  number={9},
  pages={1650--1666},
  year={2005}
}

@article{balwada2022direct,
  title={Direct observational evidence of an oceanic dual kinetic energy cascade and its seasonality},
  author={Balwada, Dhruv and Xie, Jin-Han and Marino, Raffaele and Feraco, Fabio},
  journal={Science Advances},
  volume={8},
  number={41},
  pages={eabq2566},
  year={2022},
  publisher={American Association for the Advancement of Science}
}

@article{lee1952,
  title={On some statistical properties of hydrodynamical and magneto-hydrodynamical fields},
  author={Lee, TD},
  journal={Quarterly of Applied Mathematics},
  volume={10},
  number={1},
  pages={69--74},
  year={1952}
}

@Article{onsager1949statistical,
  title={Statistical hydrodynamics},
  author={Onsager, Lars},
  journal={Il Nuovo Cimento (1943-1954)},
  volume={6},
  number={Suppl 2},
  pages={279--287},
  year={1949},
  publisher={Societ{\`a} Italiana di Fisica Bologna}
}

@article{cichowlas2005effective,
  title={Effective dissipation and turbulence in spectrally truncated Euler flows},
  author={Cichowlas, Cyril and Bona{\"\i}ti, Pauline and Debbasch, Fabrice and Brachet, Marc},
  journal={Physical review letters},
  volume={95},
  number={26},
  pages={264502},
  year={2005},
  publisher={APS}
}

@article{jha2025cascades,
  title={Cascades transition in generalised two-dimensional turbulence},
  author={Jha, Vibhuti Bhushan and Seshasayanan, Kannabiran and Dallas, Vassilios},
  journal={Journal of Fluid Mechanics},
  volume={1008},
  pages={A23},
  year={2025},
  publisher={Cambridge University Press}
}

@article{bouchet2012statistical,
  title={Statistical mechanics of two-dimensional and geophysical flows},
  author={Bouchet, Freddy and Venaille, Antoine},
  journal={Physics reports},
  volume={515},
  number={5},
  pages={227--295},
  year={2012},
  publisher={Elsevier}
}

@article{krstulovic2009cascades,
  title={Cascades, thermalization, and eddy viscosity in helical Galerkin truncated Euler flows},
  author={Krstulovic, G and Mininni, Pablo Daniel and Brachet, ME and Pouquet, A},
  journal={Physical Review E—Statistical, Nonlinear, and Soft Matter Physics},
  volume={79},
  number={5},
  pages={056304},
  year={2009},
  publisher={APS}
}

@article{onorato2015route,
  title={Route to thermalization in the $\alpha$-Fermi--Pasta--Ulam system},
  author={Onorato, Miguel and Vozella, Lara and Proment, Davide and Lvov, Yuri V},
  journal={Proceedings of the National Academy of Sciences},
  volume={112},
  number={14},
  pages={4208--4213},
  year={2015},
  publisher={National Academy of Sciences}
}

@article{miller1992statistical,
  title={Statistical mechanics, Euler’s equation, and Jupiter’s Red Spot},
  author={Miller, Jonathan and Weichman, Peter B and Cross, MC},
  journal={Physical Review A},
  volume={45},
  number={4},
  pages={2328},
  year={1992},
  publisher={APS}
}

@article{Dallasetal2015,
  title={Statistical equilibria of large scales in dissipative hydrodynamic turbulence},
  author={Dallas, Vassilios and Fauve, Stephan and Alexakis, Alexandros},
  journal={Physical review letters},
  volume={115},
  number={20},
  pages={204501},
  year={2015},
  publisher={APS}
}

@article{Agouaetal2025,
  title={Coexistence of two equilibrium configurations in two-dimensional turbulence},
  author={Agoua, Wesley and Yin, Xi-Yuan and Wu, Tong and Bos, Wouter JT},
  journal={Physical Review Fluids},
  volume={10},
  number={3},
  pages={034604},
  year={2025},
  publisher={APS}
}

@article{FoxOrszag1973,
  title={Inviscid dynamics of two-dimensional turbulence},
  author={Fox, Douglas G and Orszag, Steven A},
  journal={The Physics of Fluids},
  volume={16},
  number={2},
  pages={169--171},
  year={1973},
  publisher={American Institute of Physics}
}

@article{kraichnan1973helical,
  title={Helical turbulence and absolute equilibrium},
  author={Kraichnan, Robert H},
  journal={Journal of Fluid Mechanics},
  volume={59},
  number={4},
  pages={745--752},
  year={1973},
  publisher={Cambridge University Press}
}

\end{document}